\documentclass[a4paper,fleqn]{cas-dc}
\usepackage[authoryear]{natbib}
\usepackage{aas_macros}
\usepackage[switch]{lineno} 
\def\tsc#1{\csdef{#1}{\textsc{\lowercase{#1}}\xspace}}
\tsc{WGM}
\tsc{QE}
\tsc{EP}
\tsc{PMS}
\tsc{BEC}
\tsc{DE}

\hypersetup{
	bookmarks=true,         
	unicode=false,          
	pdftoolbar=true,        
	pdfmenubar=true,        
	pdffitwindow=false,     
	pdfstartview={FitH},    
	pdftitle={Nitric_oxide_Mars},    
	pdfauthor={Raghuram},     
	pdfsubject={Subject},   
	pdfcreator={Latex},   
	pdfproducer={Texstudio}, 
	pdfnewwindow=true,      
	colorlinks=true,       
	linkcolor=blue,          
	citecolor=blue,        
	filecolor=cyan,         
	urlcolor=blue       
}

\usepackage{setspace}

\begin{document}
\let\WriteBookmarks\relax
\def\floatpagepagefraction{1}
\def\textpagefraction{.001}
\shorttitle{Nitric oxide in the dayside of Martian upper atmosphere}
\shortauthors{Raghuram, Bhardwaj, and Dharwan}

\title [mode = title]{Model for 
	Nitric oxide and its dayglow emission in the Martian upper
	atmosphere using NGIMS/MAVEN 
	measured neutral and ion densities}                      



\author[1,2,3]{Susarla Raghuram}[type=editor,
                        auid=000,bioid=1,
                        prefix=,
                        orcid=0000-0002-1309-2499]
\cormark[1]
\ead{raghuramsusarla@gmail.com}

\credit{Conceptualization, Methodology, Software and Interpretation}

\address[1]{Physical Research Laboratory,  Ahmedabad, 380009,  India.}

\address[2]{Laboratory for Atmospheric and Space Physics, University of Colorado Boulder, Boulder, CO, USA.}

\address[3]{Space and Planetary Science Center, Khalifa University of Science and Technology, Abu Dhabi, UAE.}

\address[4]{Maulana Azad National Insitute of Technology Bhopal, M. P., India.}

\author[1]{Anil Bhardwaj}[orcid=0000-0003-1693-453X]

\author[1,4]{Maneesha Dharwan}[orcid =0000-0001-6917-4963]





\cortext[cor1]{Corresponding author}


\begin{abstract}
A comprehensive study of Nitric oxide (NO) chemistry in 
the Martian upper atmosphere
is restricted due to the lack of requisite measurements. 
 {NO is an  abundant form of odd nitrogen species 
in the  Martian lower atmosphere and its density depends on several photochemical 
processes.}
We have developed a photochemical model to study
the NO density in the dayside of 
Martian upper atmosphere by accounting for various production and loss
mechanisms. By utilizing the Neutral Gas and Ion Mass
Spectrometer (NGIMS) on-board Mars Atmosphere and 
Volatile Evolution (MAVEN) mission measured neutral and ion
densities during  deep 
dip 8 and 9 campaigns, {we modelled NO number density in
the Martian sunlit upper atmosphere for the altitudes between 120 and
200 km.} 
The modelled NO densities are employed to 
calculate  NO (1,0) $\gamma$ band emission intensity profiles in 
the dayside upper atmosphere of Mars. The calculated NO
density and its $\gamma$ band intensity profiles are found to 
be consistent with Imaging Ultraviolet Spectrograph (IUVS) onboard MAVEN 
observations and also with other modelling studies.  
We 
found that the local CO$_2$ and N$_2$ density 
variations can lead to a change in NO  density and consequently its
dayglow intensity by a factor of 2 to 5. 
Since NO is a trace constituent and also its dayglow emissions are 
strongly obscured by CO Cameron band emissions, we suggest that the 
derivation of NO number density based on our approach can constrain its 
abundance in the dayside upper atmosphere of Mars. More observations of 
(1-0) $\gamma$ band emission along 
with modelling will help to study  {the} global distribution of NO in the Martian 
atmosphere.

\end{abstract}


%
%
%


\maketitle

\section{Introduction}

 {Many in-situ and
remote observations along with theoretical studies} were carried out to 
understand the composition and energetics of the Martian 
upper atmosphere 
\citep{Nier77, Krasnopolsky93, Bertaux05, 
	Withers06, Gonzalez09, Bougher15a, Mahaffy15, 
	Bhardwaj16, Bhardwaj17}. However, the study of minor
neutrals in 
Martian upper atmosphere is still sparse due to  lack of
several key chemical reaction parameters and insufficient 
measurements from  remote and in-situ observations.
Nitric oxide (hereafter NO) is  a minor species and its density in the dayside
of Martian upper atmosphere is not well constrained. 
{Several studies have shown that though the abundance of 
this species  is in parts per 
million by volume in the thermosphere of Earth, but can 
play a significant role in driving the chemistry and thermal 
balance of the upper atmosphere \citep{Barth64, Norton70,
	Eparvier92, Barth03, Bharti19}. 
Due to its  lowest ionization potential (9.26 eV) compared to any other  atmospheric species,  NO 
quickly does exchange the charge with O$_2^+$ and causes  NO$^+$ as the dominant ion at the lower altitudes around 80 km in the Martian ionosphere \citep{Fox04}.
Hence, this species plays  a significant role in the thermospheric energy budget and also  in determining the 
ionospheric composition in the Martian atmosphere.} 
But the study of NO global distribution and 
its role in the energy budget of Martian thermosphere is limited due to the lack of 
adequate in-situ and remote observations.

 {The global distribution of NO in the Martian upper atmosphere 
	can be studied  from the remote observations of its
	 $\gamma$  band emissions in the ultraviolet region. 
	But the observation of these band emissions 
	and the derivation of NO density in sunlit 
	Martian upper atmosphere  is challenging  due to the presence of intense CO 
	Cameron bands in the  same emitting wavelength region.
	Spectroscopy
for the Investigation of the Characteristics of the
Atmosphere of Mars (SPICAM) experiment on-board  Mars Express
(MEx) spacecraft (SPICAM/MEx) and 
Imaging Ultraviolet Spectrograph (IUVS) on-board Mars
Atmosphere and Volatile EvolutioN (MAVEN) satellite
(IUVS/MAVEN) {have} played a key role in studying the NO distribution in the night side 
of Martian upper atmosphere via  $\gamma$ (225--270 nm) and $\delta$ (190--240 nm)
 band emission observations \citep{Bertaux05a,Bertaux05,Cox08,Gagne13,Stevens19}}.  
 {Recently, 
\cite{Stevens19} reported first-ever NO
dayglow (1,0) $\gamma$ band emission (A$^2\Sigma^+$-X$^2\Pi$) 
in the  IUVS/MAVEN spectra, which were observed during 6--8 April 2016 
at solar zenith angle (SZA) 75$^\circ$}. 
The measured NO (1-0) $\gamma$ band limb radiance profiles 
have been used to retrieve NO density in the
altitude range 80 to 120 km. Above 120 km, the retrieval
process becomes tedious due to higher intensity of CO Cameron band emission
which is about 50 times brighter than NO  $\gamma$ band emission.

Regarding in-situ measurements, Upper Atmosphere Mass Spectrometer (UAMS) 
instruments
on-board Viking 1 and 2 landers made first-ever in-situ 
NO number density measurements in the dayside Martian upper
atmosphere on 20 July 1976 (solar longitude, Ls = 97$^o$) and 3
September 1976 (Ls = 118$^o$), respectively, at  three to four  
different altitudes between  110 and 150  km. 
\citep{Mcelroy76, Nier77}.   {By accounting for the important photochemical reaction 
	network, \cite{Fox93} 
modelled NO number 
density and found that their calcualted density  is smaller by a factor of 2--3 compared to the 
Viking measurement at an altitude around 115 km}.  {The} reason for the 
discrepancy between the \cite{Fox93} 
modelled and Viking observed NO densities has not been understood and remained 
as a long-standing issue due to
the lack of adequate measurements. After 
nearly four decades, since  {the} Viking 
landers 
	experiments, 
the Neutral Gas and Ion Mass Spectrometer (NGIMS) 
on-board MAVEN mission (NGIMS/MAVEN)   
measured NO number 
densities in the Martian upper atmosphere \citep{Mahaffy15,Vogt17}. However,
the NGIMS/MAVEN measured NO densities 
are an order of magnitude higher than those predicted by earlier theoretical 
models. As pointed by \cite{Stevens19}, the contamination in the measurement, 
which arises due to 
the recombination of 
atomic nitrogen and atomic oxygen within 
the instrument,  could be a reason for  {the} higher reported NO densities.

As discussed before, there are  {several} constrains to study NO number density
in the dayside Martian upper atmosphere for the both in-situ (Viking 1, 2, and  
NGIMS/MAVEN) and remote (SPICAM/MEx and IUVS/MAVEN) observations. 
 {In this paper, we present an approach of calculating the  
NO number density  in the dayside Martian upper atmosphere by incorporating 
NGIMS/MAVEN measured neutral (CO$_2$, N$_2$, and O) and ion 
(NO$^+$, O$_2^+$ and CO$_2^+$) number 
densities in the photochemical network}. We show that our approach can be used to 
constrain the NO density in the altitude range between 120 and 200 km 
based on the MAVEN measurements during deep dip 8 
and 9 campaigns. We have also 
calculated  dayside NO (1-0) $\gamma$ limb intensity profiles 
 from  {our} modelled 
NO densities and compared  {them} with  {the}
IUVS/MAVEN observations.
The methodology adopted in  {this} work is explained in 
Section \ref{sec:method}. The results obtained   
analysis are presented in Section \ref{sec:res_dis}. 
We have discussed various factors which can influence
our modelled NO density profiles in Section~\ref{sec:discus}.
This work has been summarized and conclusions are
drawn in Section~\ref{sec:sum_con}.

\section{Methodology}
\label{sec:method}
The neutral and ion number densities are taken from the
dayside NGIMS/MAVEN inbound
measurements for SZA smaller than 60$^\circ$ during deep dip 8 and 9 
campaigns. To 
extract the neutral and ion density profiles, we 
utilized level 2 data (L2), version 7 (8), revision 
3 (1) data for deep dip 8 (9)
campaign. Additional details of the L2 data product are 
available in \cite{Benna18} and the data can be accessed
from a web link 
\href{https://pds-atmospheres.nmsu.edu}{(https://pds-atmospheres.nmsu.edu)}. 
We chose only those orbits of MAVEN deep dip 8 and 9 campaigns where both
neutral and ion densities are measured. The obtained densities are 
interpolated  over a uniform grid of 1 km from  {MAVEN's} periapsis 
altitude to 200 km. Table~\ref{tab:dd89} summarizes the MAVEN
observational conditions, such as observation period, 
orbit numbers, Ls, and the variation 
in SZA and latitude during orbits, during deep dip 8 and 
9 campaigns.

The chemical network considered in this work is based on the 
compilation of \cite{Fox01} with updated rate coefficients from 
\cite{Mcelroy13}. These chemical reactions are tabulated  in
Table~\ref{tab:rate_coef}. The electron temperature, 
which determines the dissociative recombination rate of ions, is taken from 
\cite{Ergun15}. Using the Analytical yield spectrum (AYS) approach, we 
calculated the suprathermal electron flux and electron impact dissociation rate 
of N$_2$ in the Martian upper atmosphere. More details of degradation of solar 
flux and the
calculation of suprathermal electron flux using AYS method can be found in 
our 
earlier work 
\citep{Bhardwaj90, Bhardwaj96, Bhardwaj99a, Jain11, Jain12, 
	Jain13, Raghuram20a}.  By assuming photochemical
equilibrium condition and  accounting for various production and loss 
mechanisms, 
we determined the NO, N($^4$S) and N($^2$D) number densities 
for the altitudes between 120 and 200 km. 
 {The effect of transport in  calculating the number densities of these species 
is discussed  {in Section~\ref{sec:discus}}.}

We have taken the fluorescence efficiency (g-factor) for NO (1,0) $\gamma$ band emission
as 2.68 $\times$ 10$^{-6}$ photons s$^{-1}$ molecule$^{-1}$  and all the photorates 
are scaled to heliocentric distance of 1.57 AU \citep{Stevens19}. 
 The Volume Emission Rate
(VER), which is the number of photons
scattered per unit volume per unit time, is calculated by
multiplying the modelled NO number density with  {the}
g-factor. The limb emission intensity of
this band  is obtained by integrating the 
volume emission rate along the line of sight 
and converted into brightness in Rayleigh (1 Rayleigh =
10$^6$/4$\pi$ photons cm$^{-2}$ sec$^{-1}$ sr$^{-1}$) using the
following equation.

\begin{equation}
\label{eq:limb}
I = 2 \times 10^{-6} \int_0^\infty 
VER(r)\ dr 
\end{equation}

here r is the abscissa along the horizontal line 
of sight and VER(r) is the volume emission rate 
(photons cm$^{-3}$ sec$^{-1}$) at a particular emission
point r. The factor 2 multiplication in the above equation is due to 
the symmetry along the line of sight concerning the
tangent point. While calculating the limb intensity, we
assumed that the emission rate is constant along the local
longitude/latitude. Other Martian species cannot absorb 
NO emissions along the line of sight due to low 
absorption cross section, hence the atmospheric absorption for
this band emission can be neglected.

\begin{table}
	\caption{The observational conditions of  MAVEN deep 
		dip 8 and 9 campaigns.} 
	\centering
	\resizebox{\columnwidth}{!}{
		\begin{tabular}{clccccccc} 
			\hline	
			\#DD$^1$ & {Observation period} & {Orbits} & SZA$^2$ & 
			Lat.$^3$ & Ls$^4$\\ 
			\hline
			8 & 16--22 October 2017  & 5909--5947&	22--40$^\circ$ 
			& {10$^\circ$N--21$^\circ$N}  & {76$^\circ$}\\
			9 & 24--30 April 2018 & 6936--6973 &  28--59$^\circ$ 
			& {19$^\circ$S--52$^\circ$S}  & 	{165$^\circ$}\\
			\hline			
	\end{tabular}}
	\footnote{Footnote}Deep dip; 			
	\footnote{Footnote}Solar zenith angle; 
	\footnote{Footnote}Latitude; \footnote{Footnote}Solar 
	longitude.
	\label{tab:dd89}
\end{table}

\begin{table*}
	\caption{Production and loss reactions of NO,
		N($^4$S) and N($^2$D) for the altitudes above 120 km 
		in the Martian upper atmosphere.} 
	\centering
	\begin{tabular}{llllll} 
		\hline
		No. & \multicolumn{2}{l}{Reaction}  & Rate coefficient & Reference \\ 	
		& & &	(cm$^3$ sec$^{-1}$ or sec$^{-1}$) & \\
		\hline
		R1 & N($^2$D) + CO$_2$ & $\rightarrow$ NO + CO & 
		3.6$\times$10$^{-13}$ & \cite{Herron99}\\	
		R2 & N($^2$D) + O & $\rightarrow$  N($^4$S) + O & 
		{6.90$\times$10$^{-13}$}& \cite{Fell90}\\	
		R3 & N($^2$D) + N$_2$ & $\rightarrow$  N($^4$S) + N$_2$ & 
		{1.70$\times$10$^{-14}$} & \cite{Herron99}\\
		R4 & N($^2$D) + NO & $\rightarrow$  N($^4$S) + NO &  
		{6.70$\times$10$^{-11}$} & \cite{Fell90}\\	
		R5 & N($^2$D) + O$_2^+$ & $\rightarrow$  NO$^+$ + O & 
		{1.80$\times$10$^{-10}$} & \cite{Goldan66}\\
		R6 & N($^2$D) + e$_{th}$ & $\rightarrow$  N($^4$S) + O & 
		{3.86$\times$10$^{-10}$} (T$_e$/300)$^{0.81}$ & 
		\cite{Berrington81}\\
		R7 & N($^4$S) + CO$_2$  & $\rightarrow$  NO + CO & 
		{1.7$\times$10$^{-16}$} & \cite{Fox01}\\ 	 
		R8 & N($^4$S) + NO  & $\rightarrow$  N$_2$ + O & 
		3.38$\times$10$^{-11}$ (300/T$_n$)$^{0.17}$ exp(-2.8/T$_n$) & 
		\cite{Mcelroy13}\\		
		R9 &N($^4$S) + CO$_2^+$ & $\rightarrow$  NO +CO$^+$  
		& 3.40$\times$10$^{-10}$ & \cite{Scott98}\\
		R10 &N($^4$S) + O$_2^+$ & $\rightarrow$  NO$^+$ + O & 
		1.80$\times$10$^{-10}$& \cite{Mcelroy13}\\
		
		R11 & NO$^+$ + e$_{th}$ & $\rightarrow$  N($^4$S)+ O & 
		0.60$\times$10$^{-7}$(300/Te)$^{0.5}$ & 
		\cite{Vejby98}\\
		R12 & NO$^+$ + e$_{th}$ & $\rightarrow$  N($^2$D) + O & 
		3.40$\times$10$^{-7}$(300/Te)$^{0.5}$ & 
		\cite{Vejby98}\\
		R13 & NO + CO$_2^+$ & $\rightarrow$  NO$^+$ + CO$_2$ &
		1.2 $\times$10$^{-10}$ & \cite{Mcelroy13}\\ 
		R14 & NO + O$_2^+$ & $\rightarrow$  NO$^+$ + O$_2$ &
		4.6 $\times$10$^{-10}$ & \cite{Mcelroy13}\\        
		R15 & N$_2$ + h$\nu$ & $\rightarrow$  N($^4$S) + N($^4$S) & 
		Calculated & This work\\		
		R16 & N$_2$ + h$\nu$ & $\rightarrow$  N($^2$D) + N($^4$S) &
		Calculated & This work\\   
		R17 & N$_2$ + e$_{ph}$ & $\rightarrow$  N($^4$S) + N($^4$S) & 
		Calculated & This work\\		
		R18 & N$_2$ + e$_{ph}$ & $\rightarrow$  N($^2$D) + N($^4$S) &
		Calculated & This work\\  
		R19 & N$_2$ + O$_2^+$ & $\rightarrow$  NO$^+$ + NO &
		 {2.0 $\times$10$^{-18}$} &  {\cite{Matsuoka81}}\\ 
		R20 & N($^2$D)  & $\rightarrow$  N($^4$S) + h$\nu$ &
		2.78 $\times$10$^{-5}$ & \cite{Tachiev02}\\    
		\hline
	\end{tabular}
	\label{tab:rate_coef}
	
	{h$\nu$ is photon, e$_{ph}$ is photoelectron and e$_{th}$ is thermal 
		electron; T$_n$ and T$_e$ are neutral and electron temperatures, 
		respectively..}
\end{table*}

\section{Results} 
\label{sec:res_dis}
\begin{figure}
	\includegraphics[width=\linewidth]{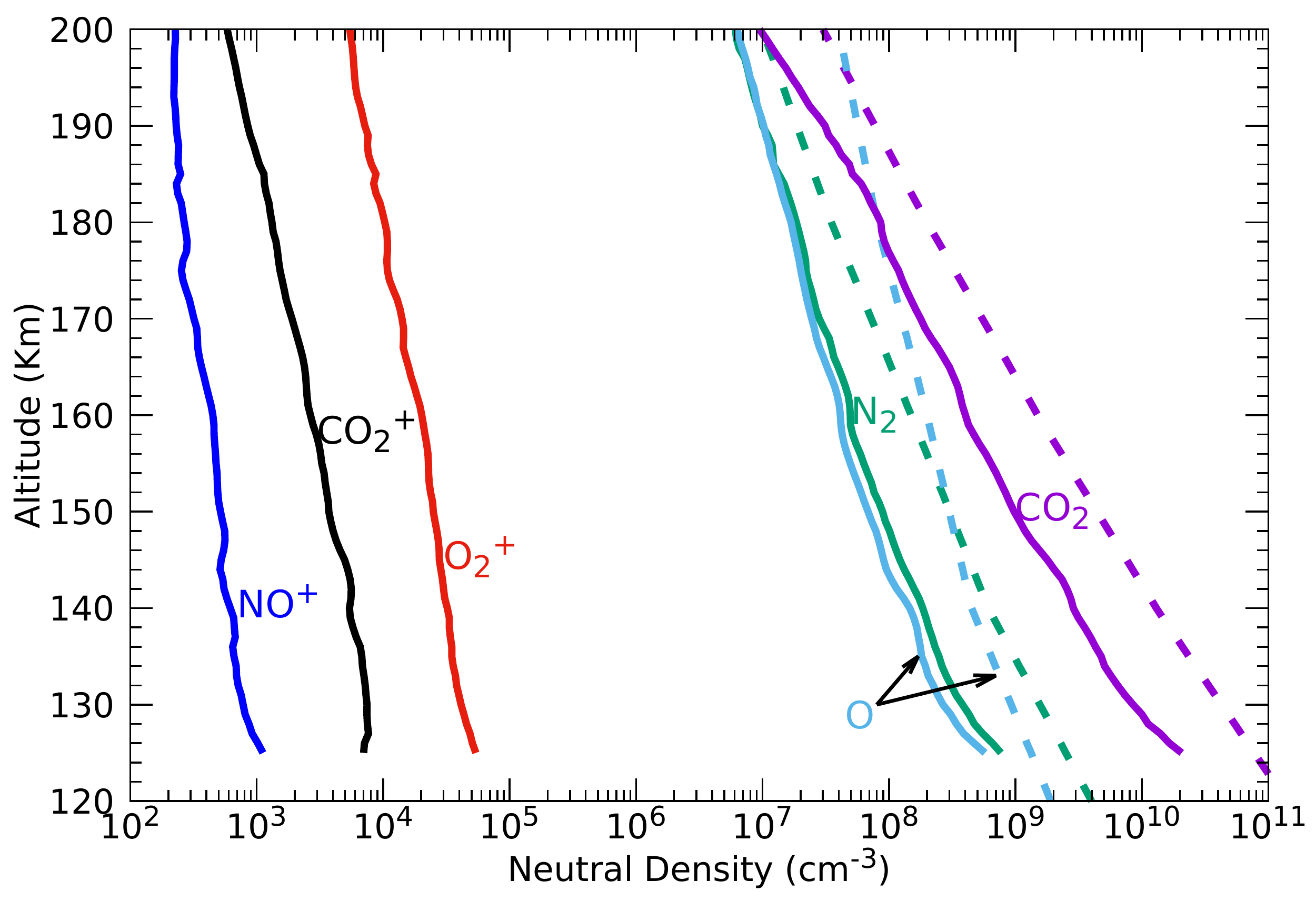}
	\caption{The NGIMS/MAVEN measured neutral and ion 
		number density profiles for  orbit \#5947 of
		deep dip 8 campaign. 
		The  dashed curves with corresponding colours
	    represent the neutral
	    density profiles for solar minimum condition from \cite{Fox04}.}
	\label{fig:dens_ionsneu}
\end{figure}

The NGIMS/MAVEN measured neutral 
and ions densities profiles for orbit \#5947 of deep dip 8 campaign  
are
presented in Figure~\ref{fig:dens_ionsneu}. In this Figure, we compared {the} 
NGIMS/MAVEN measured  
CO$_2$, N$_2$ and O density profiles with \cite{Fox04} modelled density 
profiles, which are constructed based on  {the} Viking observations 
for solar minimum condition.  It can be noticed in this figure that \cite{Fox04} modelled neutral density profiles are 
higher by a  factor of  {2 to 4 compared to NGIMS/MAVEN measured values.}

\begin{figure*}
	\centering
	\includegraphics[width=\linewidth]{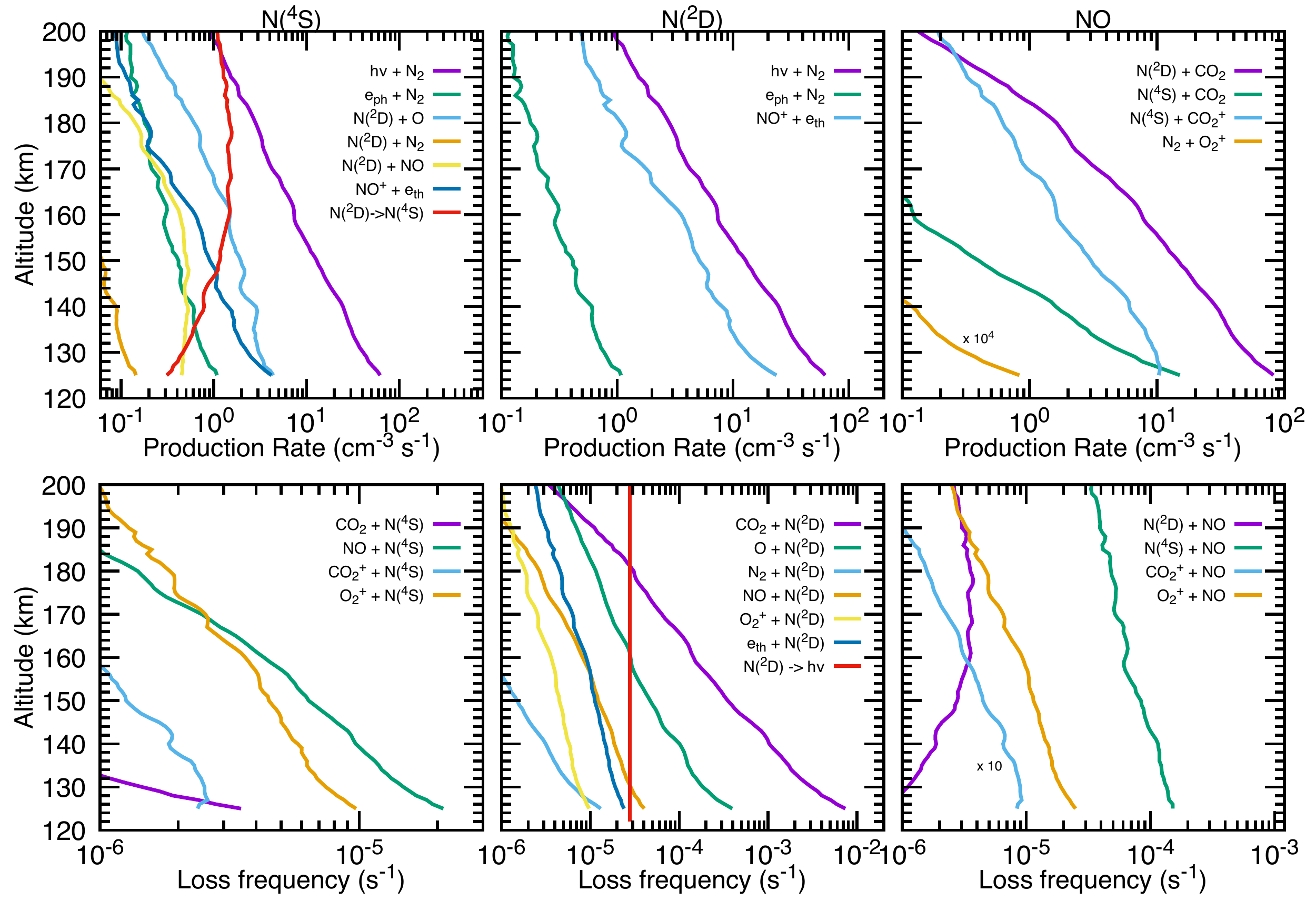}
	
	\caption{Various modelled formation (top panels) and destruction 
		(bottom panels) processes for N($^4$S), N($^2$D) and NO   in the 
		Martian upper 
		atmosphere for  orbit \#5947 during MAVEN deep dip 
		8 campaign. 
		h$\nu$,
		e$_{ph}$, and e$_{th}$ represent the solar photon, 
		photoelectron, and thermal electron, respectively. Production rate of 
		NO via N$_2$ and O$_2^+$ is plotted after multiplying by a factor of 
		10$^4$. The 
		reaction profile CO$_2^+$ and NO, which is a loss frequency profile of 
		NO, is 
		plotted after multiplying by a factor of 10.}
	\label{fig:pl_nnno}
\end{figure*}

 {Using the NGIMS/MAVEN measurements for  orbit \#5947,}  the modelled production and loss profiles of N($^4$S), 
N($^2$D), and NO are presented in the top and bottom panels of 
Figure~\ref{fig:pl_nnno}, respectively. The calculated production rate 
profiles in the top 
panels of this Figure show that photodissociation of 
N$_2$ is the major source of N($^4$S) and N($^2$D) production, whereas the 
collisional 
reaction between N($^2$D) and CO$_2$ leads to the formation of NO. 
 {For the altitudes above 180 km, radiative decay of N($^2$D)  also contributes 
significantly to the total formation of N($^4$S), whereas the collisional reaction 
between CO$_2^+$ and N($^4$S) is also {an} important source mechanism in the  total NO formation}.  It can be 
noticed in this Figure that the role of other chemical processes is  small in
the formation  of N($^4$S), N($^2$D) and NO when compared 
to  previously discussed major production mechanisms.

 {The modelled loss frequency profiles in the 
bottom panels of Figure~\ref{fig:pl_nnno} show that for the altitudes below 160 
km, the total loss of N($^4$S) 
is mainly due to  {the} collisions with NO,  and above this distance it is removed by the collisions of O$_2^+$ in the Martian upper atmosphere.}
The loss frequency of N($^4$S) by CO$_2$ collision is smaller by an order of 
magnitude or more compared to that of the previously discussed loss mechanisms 
which is
due to small collisional rate coefficient (more 
than five order of magnitude smaller, see reactions 
R7, R8, and R10 in Table~\ref{tab:rate_coef}). For the altitudes below 180 km, 
the total loss of N($^2$D) is mainly 
due to the collisions with CO$_2$, which leads to the formation of NO, and
above this altitude the radiative decay is  the dominant loss mechanism that 
causes N($^4$S) formation. The loss 
of NO is 
mainly due to the 
collisions with N($^4$S) which leads to the formation of N$_2$. 
 {Several other reactions are also involved in the loss of these 
species but their contribution to the total loss frequency is small.}

\begin{figure}
	\centering
	\includegraphics[width=\linewidth]{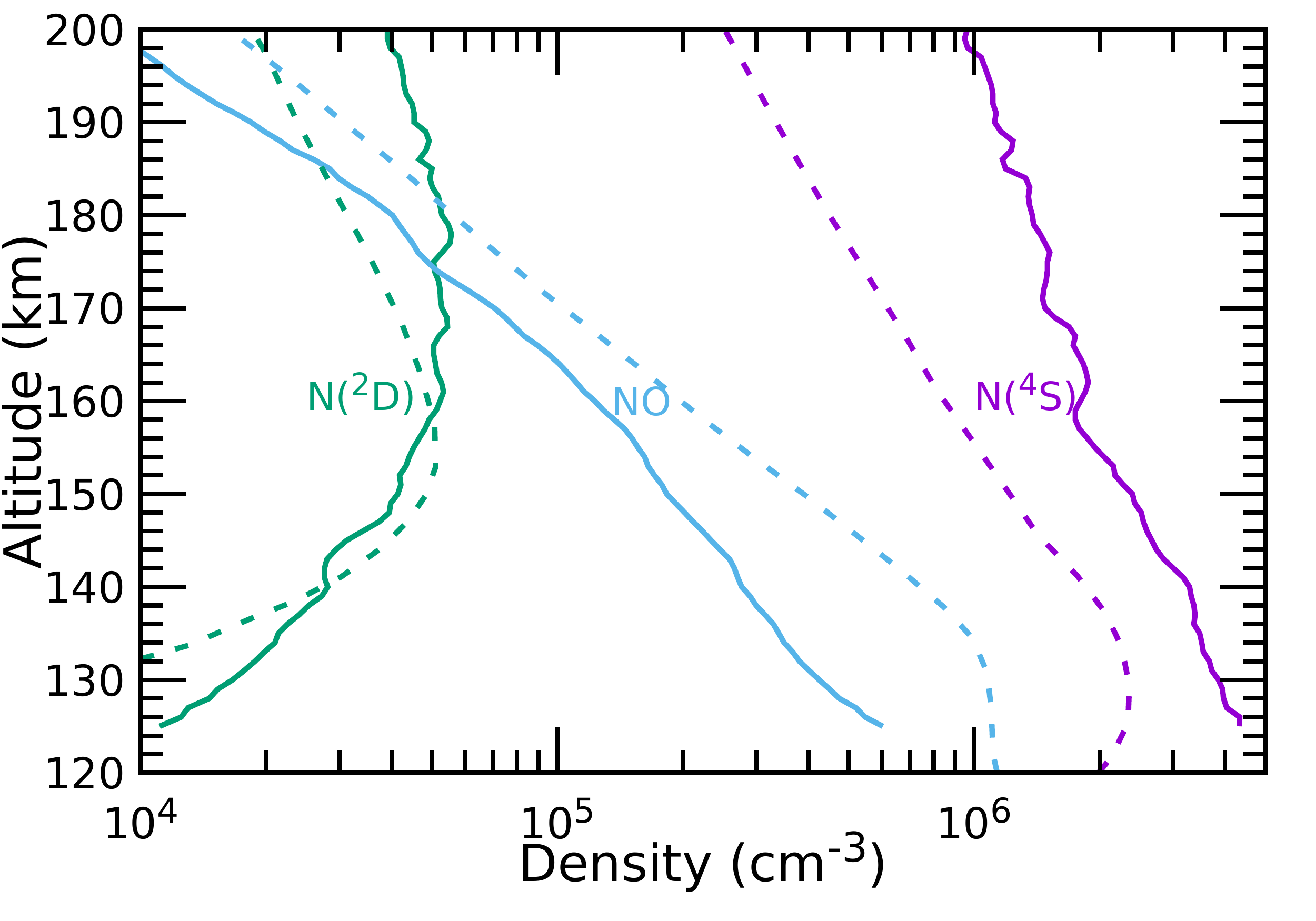}
	\caption{Modelled N($^4$S), N($^2$D), and NO densities in the Martian
		upper atmosphere  for orbit \#5947 of MAVEN deep dip 8 campaign.
	The  dashed curves with corresponding colours represent the modelled density profiles from \cite{Fox04} for solar minimum condition.}
	\label{fig:oddn_dens}
\end{figure}

 The modelled number 
density profiles of N($^4$S),  N($^2$D) and NO 
for  orbit \#5947 of MAVEN deep dip 8 campaign 
are presented in 
Figure~\ref{fig:oddn_dens}.   Our modelled NO and N($^4$S) density profiles 
are lower and higher by a factor of 2 to 3, respectively, compared those of  
\cite{Fox04}. Our modelled N($^2$D) density profile is closer to \cite{Fox04} 
calculated value 
for the altitude above 130 km.  The reason for  differences between these density profiles is discussed in Section~\ref{sec:discus}.

\begin{figure*}
	\centering
	\includegraphics[width=0.85\linewidth]{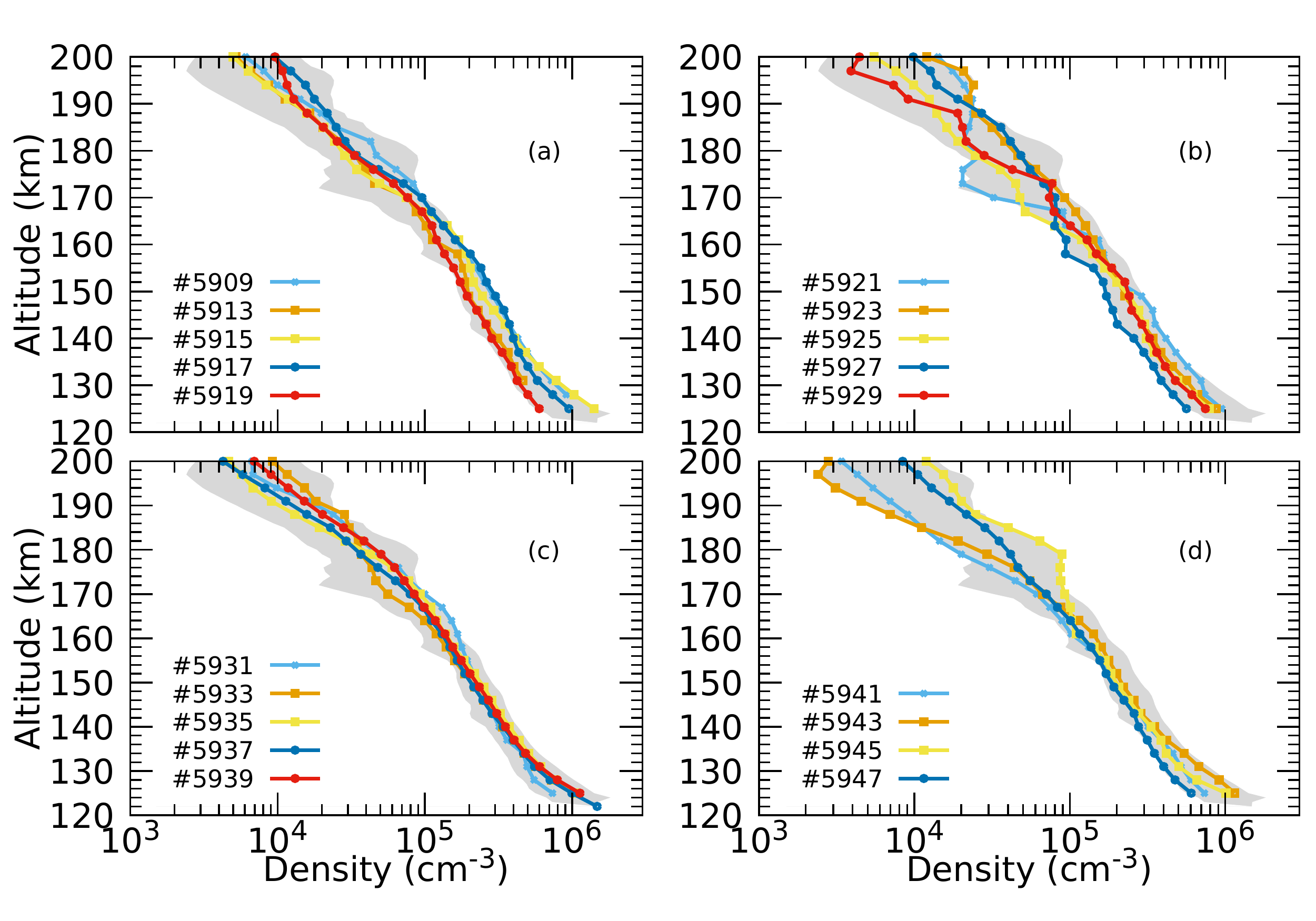}
	\caption{The modelled NO number density 		
		profiles for the various orbits of MAVEN deep dip 8 
		campaign. The Gray shaded area represents the 
		variability  in the calculated NO number density.}
	\label{fig:no_den_dd8}
\end{figure*}

\begin{figure*}
	\centering
	\includegraphics[width=\linewidth]{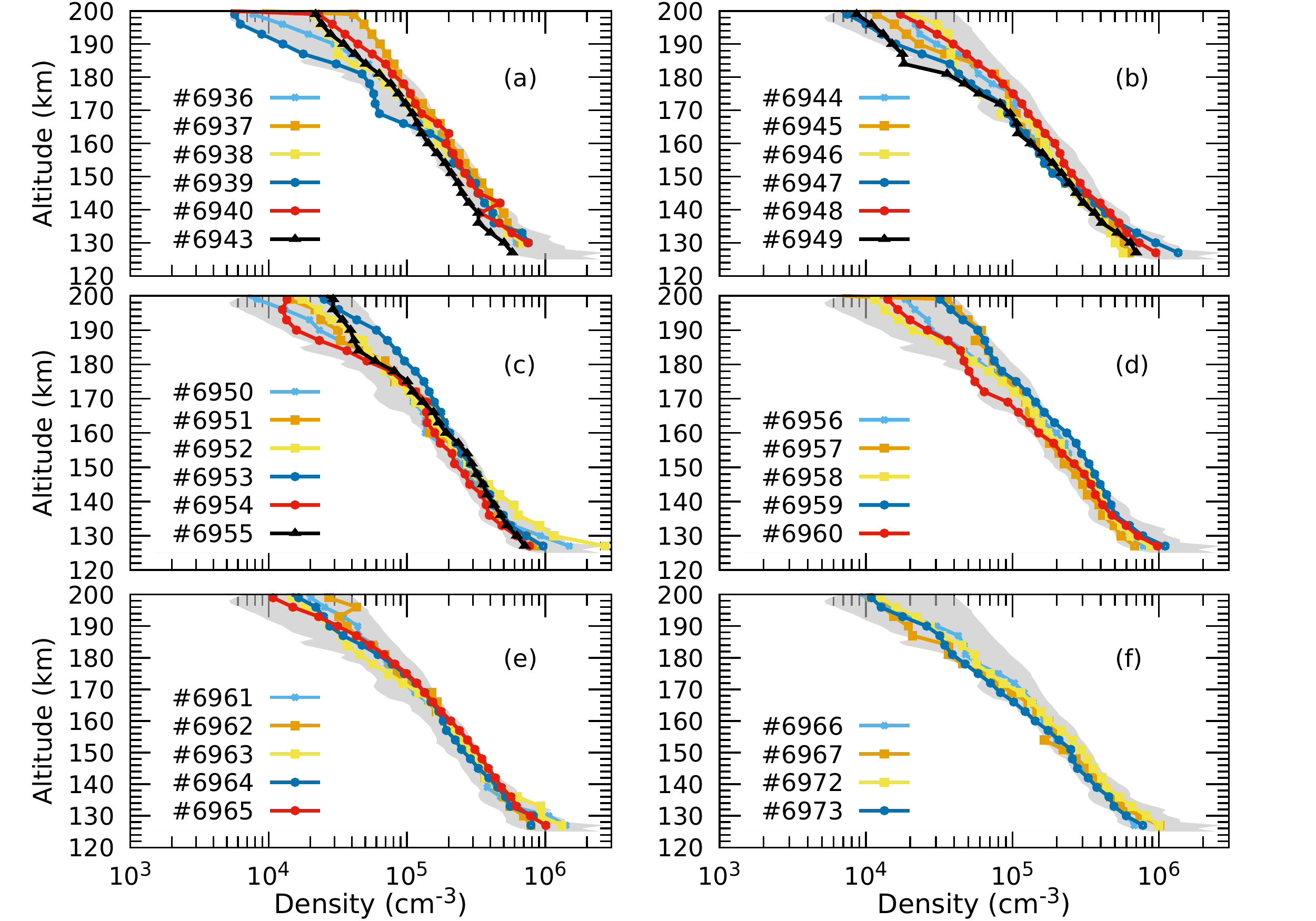}
	\caption{The calculated NO number density 
		profiles for various orbits of MAVEN during Deep dip 9 
		campaign. The Gray shaded area represents the 
		variability in the modelled NO number density.}
	\label{fig:no_den_dd9}
\end{figure*}

We have employed the same approach in calculating the NO number 
density profiles
for all the orbits of 
MAVEN deep dip 8 and 9 campaigns. 
Using the NGIMS/MAVEN  measured neutral and ion number density 
 for the  orbits  
of deep dip 8 campaign (\#5909 to \#5947)  and accounting for the previously discussed 
photochemical network of reactions, 
the modelled NO number 
density profiles  are presented in  Figure 
\ref{fig:no_den_dd8}. We find that our modelled NO number density is 
varying over different orbits and this is mainly
associated with the variability in the  NGIMS/MAVEN 
measured neutral and ions density measurements.
The variation in the calculated NO number density for different orbits
is found to be small (by a factor of 2) at lower altitudes
(around 120 km) 
and it is increasing with altitude (more than a factor of 2, see the gray 
shaded area in Figure~\ref{fig:no_den_dd8}).
The calculated NO number density profiles 
for all the 32 orbits, i.e., from \#6936 to \#6973,
of MAVEN deep dip 9 campaign are 
presented in  Figure~\ref{fig:no_den_dd9}. 
In this case, the calculated NO density around 120 km altitude is about 
10$^6$ cm$^{-3}$  
and comparable  to the derived value for the deep dip 8 campaign.

\begin{figure}
	\includegraphics[width=\linewidth]{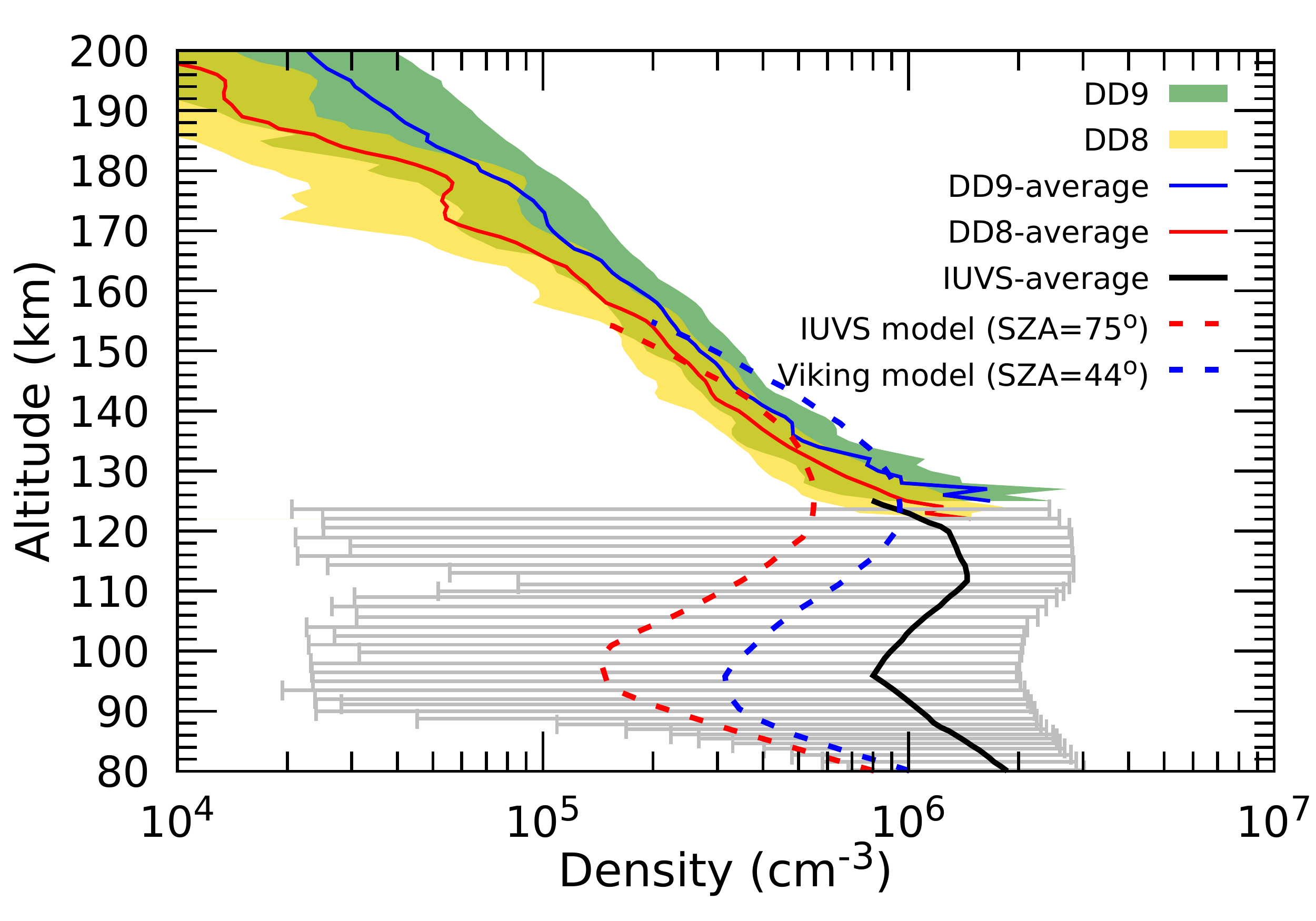}
	
	\caption{Comparison between the  
		modelled and MAVEN deep dip 8 (solid red curve for the average),  9 
		(solid 	blue curve red for the	average) and  
		IUVS/MAVEN (solid black curve) retrieved mean and Viking modelled at 
		SZA 44$^o$ 	(blue-dashed curve) and IUVS modelled at SZA 75$^o$ 
		(red-dashed curve) NO number density 
		profiles in the Martian upper atmosphere. 
		The yellow and green shaded areas 
		represents the variability in calculated NO number density 
		profiles during MAVEN deep dip 8 and 9  campaigns,
		respectively. The horizontal Gray error bars 
		represents the digitized data of 1-$\sigma$ variability
		about the mean value of the IUVS
		derived NO number density \citep[taken from][]{Stevens19}.}
	\label{fig:no_cmp}
\end{figure}

In Figure~\ref{fig:no_cmp}, we  compared the calculated NO number density 
profiles during MAVEN deep dip 8 and 9 campaigns with IUVS/MAVEN 
retrieved  value.  The 
modelled average NO number density profile for deep dip 8 campaign  is consistent with the IUVS/MAVEN  
retrieved value at around 125 km altitude. \cite{Stevens19} modelled NO density 
profiles 
in the altitude range 80 and
160 km at	 SZA 44$^\circ$ and 
75$^\circ$ for IUVS and Viking observational conditions, respectively  (see 
dashed red and blue curves in Figure~\ref{fig:no_cmp}). Our modelled  NO number 
density 
profiles are also consistent with the \cite{Stevens19} modelled 
 profiles in the altitude range 120 to 160 km.

\begin{figure}
	\centering
	\includegraphics[width=\linewidth]{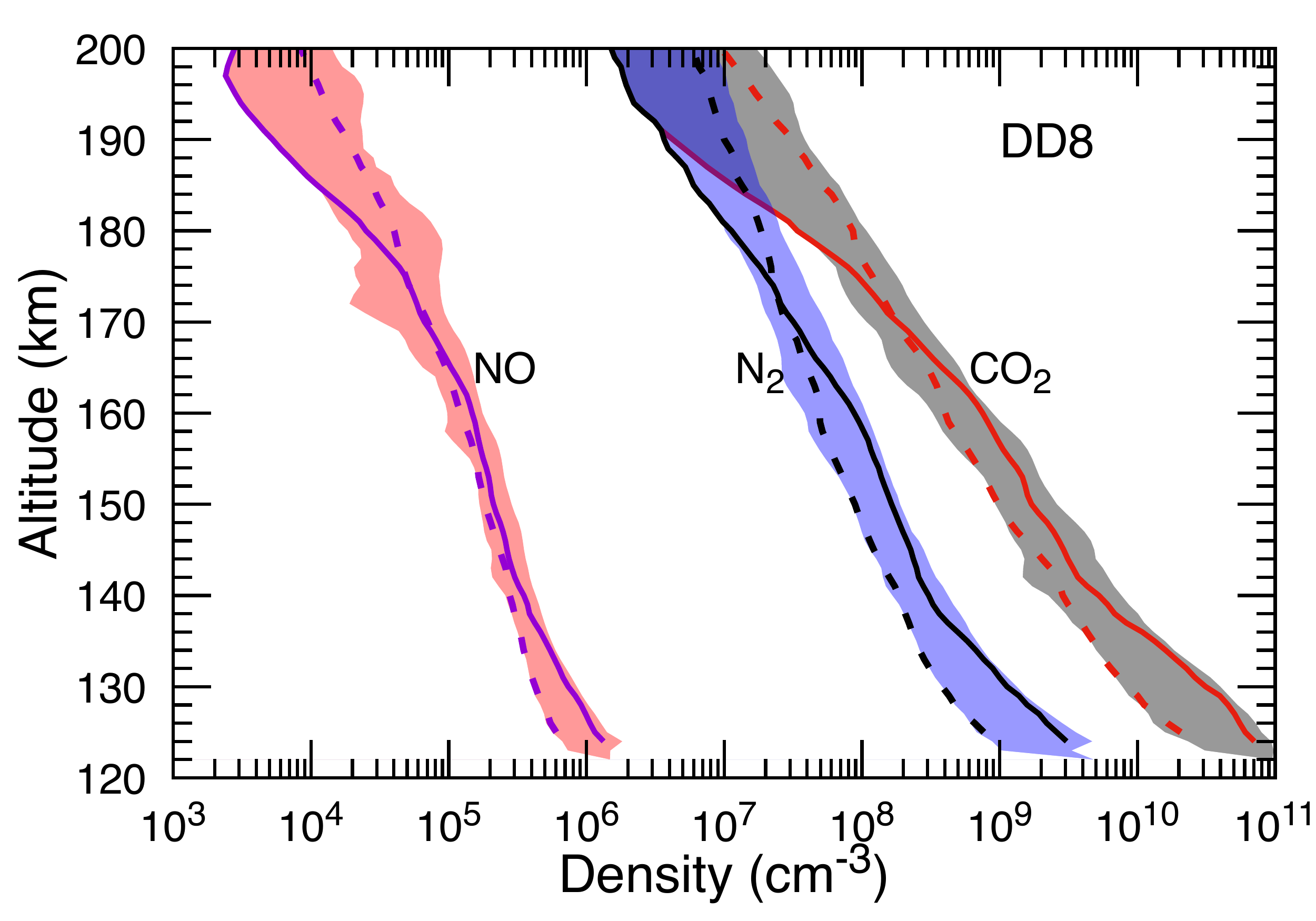}
	\includegraphics[width=\linewidth]{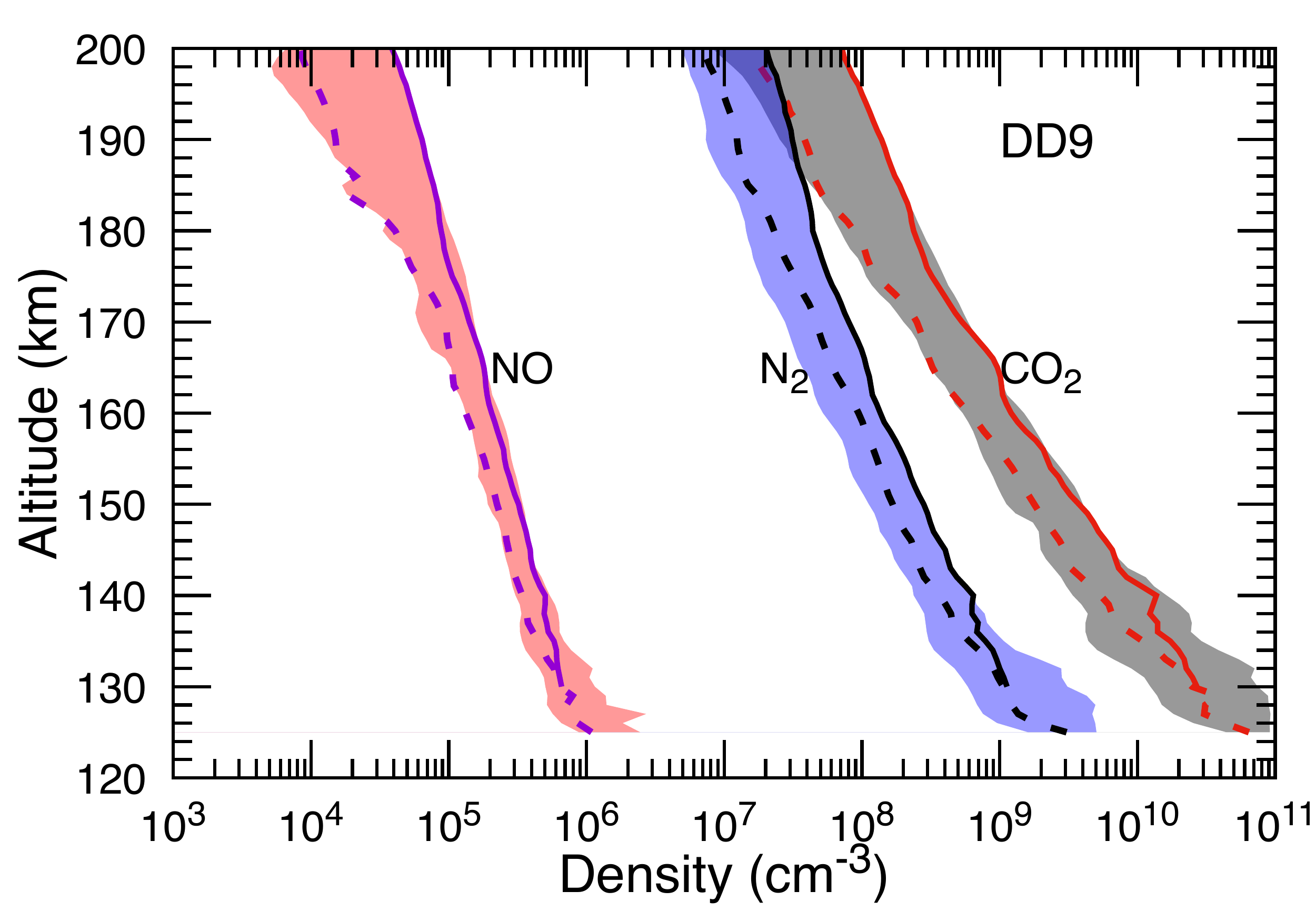}
	\caption{Variation in the modelled  NO density due to the change in the  NGIMS/MAVEN measured CO$_2$ and  N$_2$ densities  during the MAVEN deep 
		dip 8 (top panel) and 9 (top panel) campaigns. The solid and dashed curves in the top panel (bottom panel) represent the corresponding number densities for  orbits \#5943 and \#5947 (\#6937 and \#6949),  respectively, for   MAVEN
		deep dip 8 (9) campaign.} 
	\label{fig:dens_cmp}
\end{figure}

 {Our calculations in Figure~\ref{fig:pl_nnno} show that  N$_2$ 
and CO$_2$  play a crucial role in determining the  NO 
density in the Martian upper atmosphere. Hence, we studied the
variation in our modelled NO density with respect to the changes in 
MAVEN measured CO$_2$ and N$_2$ densities.   
It can be noticed in Figure~\ref{fig:dens_cmp} that both the CO$_2$
and N$_2$ densities locally vary about a factor of 2 to 5 during deep dip 8 and 
9 campaigns in the altitude range 120 to 200 km (see the shaded areas in 
 figure~\ref{fig:dens_cmp}).  Due to significant variability in the measured densities, it is difficult to describe the impact of CO$_2$ and N$_2$ on 
NO for each orbit  
of both deep dip 8 and 9 campaigns. Hence, we chose two orbits for each campaign to 
demonstrate the impact of MAVEN measurements on the modelled NO density. 
As shown in the top panel of Figure~\ref{fig:dens_cmp},  
the MAVEN measured CO$_2$ and N$_2$
densities for orbit \#5943 are higher compared to those for  \#5947  at the altitudes below 165 km
and above this radial distance it is opposite. This cross over in the measured CO$_2$ and N$_2$ density profiles is also reflected in our modelled NO profiles for the corresponding orbits 
 at 165 km altitude (see solid and dashed purple lines  in the top panel of  
 Figure~\ref{fig:dens_cmp}). Similarly, as shown in the bottom panel of this figure, the 
MAVEN measured 
CO$_2$ and N$_2$ densities for orbits \#6937 are higher compared to those for \#6949 
 for deep dip 9 campaign in the altitude range 120 to 200 km. The higher  densities of CO$_2$ and N$_2$  lead to larger production of 
 NO which is reflected in the modelled NO density for orbit \#6937. These calculations 
 suggest  that the change in  CO$_2$ and N$_2$ densities directly can influence 
 the NO density in the Martian upper atmosphere.}

\begin{figure*}
	\centering
	\includegraphics[width=\linewidth]{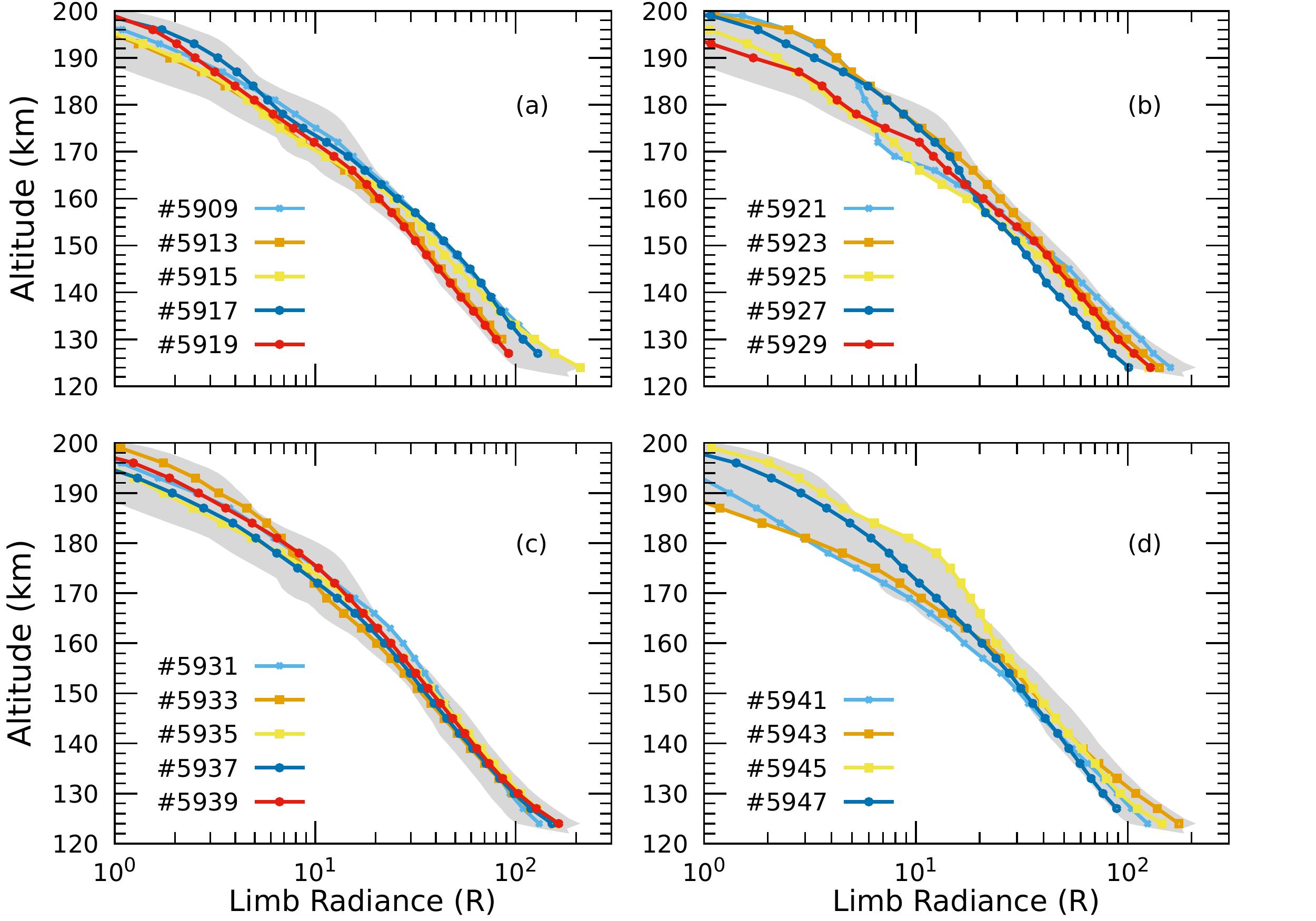}
	\caption{Modelled limb radiance profiles for Nitric oxide (NO) 1-0 
		$\gamma$ emission  for the MAVEN Deep dip 
		8 campaign. The Gray shaded area 
		represents the variability in the calculated 
		NO (1-0) $\gamma$  intensity.}
	\label{fig:limb_inten_dd8}
\end{figure*}

\begin{figure*}
	\centering
	\includegraphics[width=\linewidth]{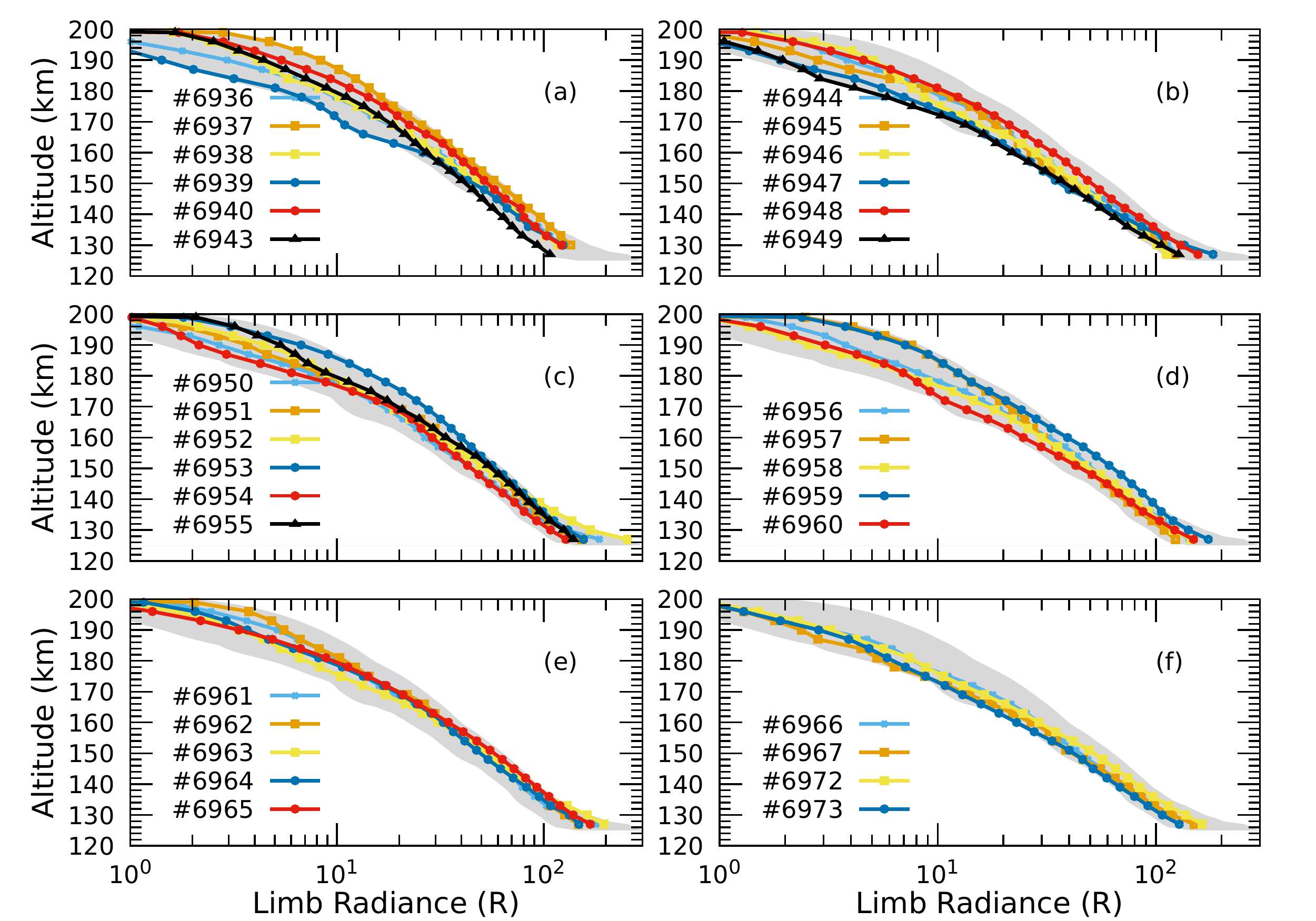}
	\caption{Modelled  limb radiance profile of Nitric oxide (NO) 
		emission   for the MAVEN deep dip 9 campaign.
		The Gray shaded area represents 
		the variability in the calculated NO (1-0) $\gamma$ 
		intensity.}
	\label{fig:limb_inten_dd9}
\end{figure*}

\begin{figure}
	\includegraphics[width=\linewidth]{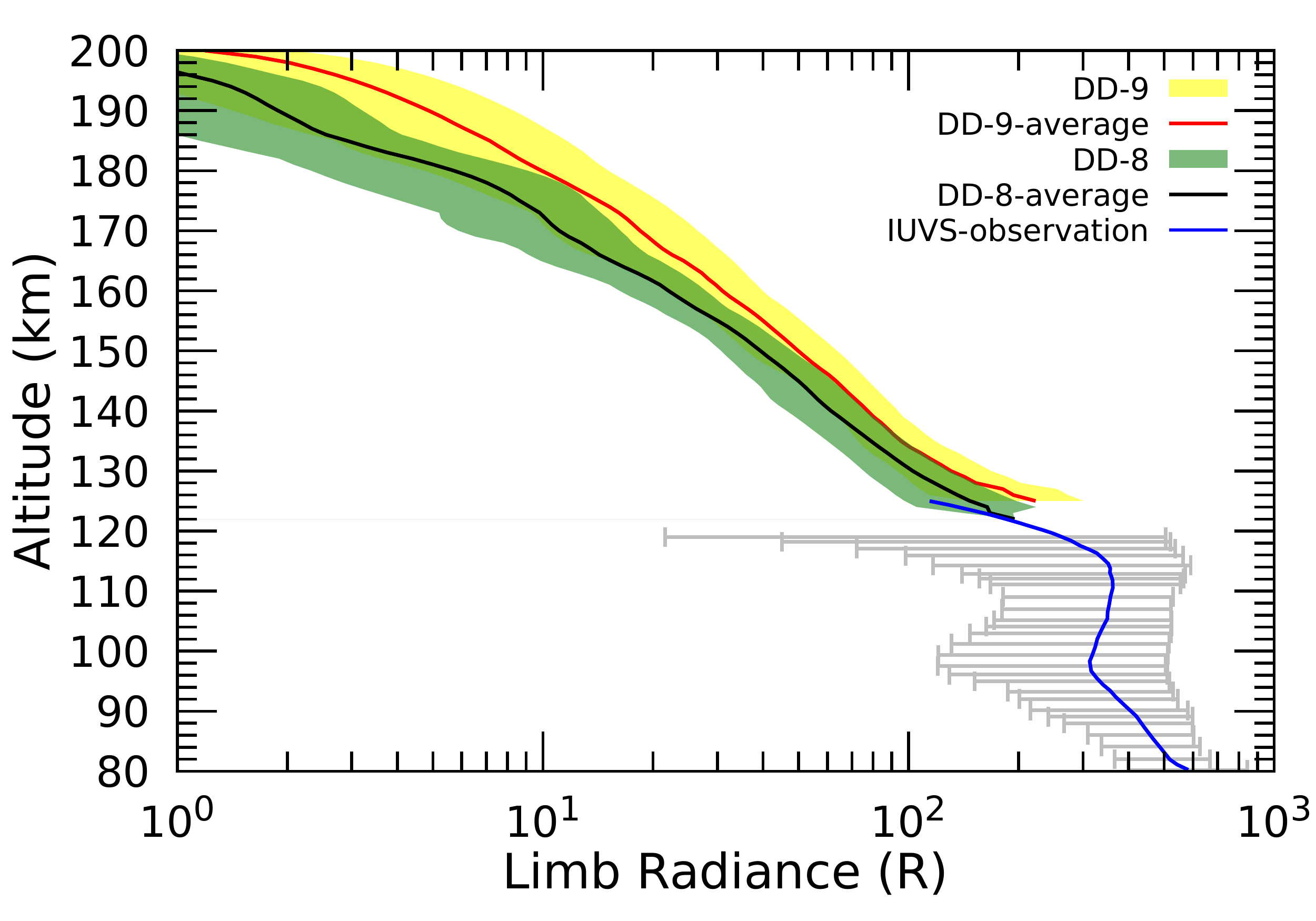}
	
	\caption{Comparison between the modelled and the 
		IUVS/MAVEN observed NO (1,0) $\gamma$ band limb intensity 
		profiles. The black and red
		curves represent the calculated average NO (1,0) $\gamma$
		band limb intensity for deep dip 8 and 9 missions
		respectively. Blue curve and gray errors bars
		represents the observed average IUVS/MAVEN dayglow 
		profile and 1-$\sigma$ uncertainty associated with
		observation, respectively \citep[taken from][]{Stevens19}. Green and yellow shaded
		areas represent the variability in the calculated 
		NO (1,0) $\gamma$ band limb intensity for deep dip 8 and 9
		missions, respectively.} 
	\label{fig:iuvs_no_cmp}
\end{figure}

By utilizing the corresponding modelled NO number densities, the calculated limb intensity profiles for NO 
(1,0) $\gamma$ band for the MAVEN deep dip 8 and 9 campaigns are presented in
Figures~\ref{fig:limb_inten_dd8} and ~\ref{fig:limb_inten_dd9}, respectively. 
The calculated NO (1,0) $\gamma$ band limb intensity is found to vary  
between 90 and 200 R (between  100
and 300 R)  at the MAVEN periapsis altitude  around 120 km during 
 MAVEN deep dip 8  (9) campaign.
We compared our calculated NO (1, 0) $\gamma$ limb intensity 
profiles with and IUVS/MAVEN observation in Figure~\ref{fig:iuvs_no_cmp}. 
Our calculated mean 
limb intensities for deep  dip 8 and 9 campaigns around 120 km altitude are 
found to be consistent with the  IUVS/MAVEN 
observation. 


\section{Discussion}
\label{sec:discus}
During the deep dip campaigns, NGIMS/MAVEN  measured both 
neutral and ion densities at the altitudes as low as 125 km,
whereas the regular measurements cover the altitudes above 
150 km. This possibility enabled us to calculate NO number
density at the lower altitudes which is not possible for other
measurements. 
Out of the total nine MAVEN deep dip campaigns so far, 
deep dip 2, 8, and 9 campaigns are occurred  {on} the dayside. 
{Recently, Cui et al. (2020)  also used the NGIMS neutral and ion density 
measurements to constrain the NO density in the dayside Martian upper atmosphere, 
specifically for the deep dip 2 campaign during 17-22 April 2015 with a periapsis 
altitude of 130 km and appropriate for the subsolar condition. By neglecting transport, 
these authors carried out time-dependent calculations at a fixed altitude of 160km to 
investigate the diurnal variation under the influences of both dayside solar radiation 
and nightside solar wind electron precipitation. }
{The present study has several advances over Cui et al. (2020) in that (1) our 
calculations are made over a broad altitude range of 120 - 200 km to provide the 
vertical distribution of NO for deep dip 8 and 9 campaigns; (2) we emphasize the dayside variability in NO density 
driven by the variability of the background atmosphere, especially in terms of the CO$_2$ 
and N$_2$ densities, rather than the mean diurnal variability focused by \cite{Cui20}; (3) we also estimate the NO (1,0) gamma band emission such that our results could 
be directly compared to the MAVEN IUVS limb observations.}
 {By  comparing our modelled NO density profiles with \cite{Cui20} calculations,
we find that  our calculated densities are smaller by a factor 2 to 3. 
This difference could be due to the different prevailed conditions in the Martian upper 
atmosphere  during deep dip 2, 8 and 9 campaigns. Moreover, \cite{Cui20}  account for 
local  time variability of neutral species in their calculations which is not the case in this work.}

{The calculations presented in Figure \ref{fig:pl_nnno}
show that the odd nitrogen species,  which are primarily  
produced from the photodissociation of N$_2$, initiate the formation of NO via 
collisions with CO$_2$, and further interaction of these species with NO  
subsequently recycles the N$_2$ in the Martian upper atmosphere (see chemical reactions R1, R8, R15 and R16 in Table~\ref{tab:rate_coef}).
So it should be noticed that   besides variation in CO$_2$ and N$_2$ densities, 
the  NO density in the Martian upper atmosphere depends significantly on chemical  
reactions particularly R1 and R8, which are respective major production and loss 
channels initiated by photodissociation of N$_2$.} However, the  
calculations presented in figure~\ref{fig:pl_nnno} suggest that the NGIMS/MAVEN 
measured major neutral and ion
densities (CO$_2$, N$_2$, and O$_2^+$) along with  a few major chemical reactions 
are sufficient to determine the 
NO density in the altitude range 120 to 200 km, rather than using a complex 
chemical network.

The difference between our calculated 
odd nitrogen species density profiles and \cite{Fox04} modelled values, which 
is about 
a factor of 2 to 5, is mainly due to the change in 
input CO$_2$ and N$_2$ densities (see Figures~\ref{fig:dens_ionsneu} and \ref{fig:oddn_dens}). But our 
modelled NO density 
profile in the altitude range 120 to 
200 km is closer to the \cite{Fox04} calculation (with a factor of 2 difference),  which 
suggests 
that  NO is strongly controlled by photochemical reactions rather 
than transport in the upper atmosphere.  Time scales of odd nitrogen species as 
determined  by \cite{Cui20} also suggest that NO is under 
photochemical equilibrium for the altitudes up to 180 km and above which 
chemical diffusion significantly controls its density \cite[see  Fig.~1d in ][]{Cui20}. Moreover, the agreement between our 
calculated NO densities for different orbits of deep dip 8 and 9 campaigns with the
\cite{Stevens19} modelled values for IUVS/MAVEN observation condition in the 
altitude range 120 to 160 km also 
supports our approach that NO can be derived from the  NGIMS/MAVEN measured 
CO$_2$ and 
N$_2$ densities and also by considering the major photochemical reactions (see 
Figure~\ref{fig:no_cmp}).

 {NO density in  the Martian upper atmosphere is significantly driven  
by photochemical reactions as well as local neutral densities of CO$_2$ and N$_2$}.  Our calculations in Figure~\ref{fig:pl_nnno} show that 
below 170 km, the chemical cycle of NO mainly depends on density distribution 
of N$_2$ and CO$_2$ in the Martian upper atmosphere. Hence, the variation in 
the measured densities of these 
species directly can impact  
the modelled NO number density. 
 {As explained before, the variability in the MAVEN measured CO$_2$ and N$_2$
densities is directly linked to the NO density in the Martian upper atmosphere  (see solid and dashed curves in Fig.~\ref{fig:dens_cmp})}
The higher volume 
mixing ratios of N$_2$ leads to the larger formation rate of 
dissociative products N($^4$S) and N($^2$D) and eventually can produce more NO. 
Since the major production source of NO is via collisional reactions between 
N($^2$D) with CO$_2$, the variation in local  CO$_2$ density can also control
the NO density. 
The ambient density variation  presented in the 
Figure~\ref{fig:dens_cmp} show that during deep dip 8 and 9 campaigns both
CO$_2$ and N$_2$ densities varied by a factor of 5 or more in the Martian upper 
atmosphere which results in the variation in our calculated NO densities.
The  calculations done by \cite{Cui20} for deep dip 
2 campaign show that the diurnal variation in  NO, N($^4$S) and N($^2$D) at the 
reference altitude 160 km is 
due to ambient N$_2$ mixing ratio at 160 km. But the calculations presented in this 
work show that besides the 
variation in  N$_2$ volume mixing ratio, the change in  CO$_2$ 
density also plays an important role in determining the NO density in the 
Martian upper atmosphere.

As shown in Figure~\ref{fig:no_cmp}, our calculated 
NO number densities 
around 125 km, for deep dip 8 and dip 9 campaigns, are
consistent with the IUVS/MAVEN measurements and also with the modelled density
profiles in the altitude range 120--160 km. It can be noticed in this figure
that at 125 km altitude,  the calculated NO number 
density varies by a factor of about 5 over the both deep dip 
campaigns, which is mainly due to the local variation in the
measured CO$_2$ and N$_2$ densities. The earlier MAVEN deep dip
campaigns also reveal that there is a variability in the measured neutrals and 
ions in the Martian upper atmosphere \citep{Bougher15}. Small scale structures in the 
NGIMS/MAVEN observed neutral and ion densities are linked to wave activity
in the upper atmosphere or solar wind interaction with ionosphere
\citep{England17,Kopf08}. \cite{Williamson19} also observed large scale 
amplitude perturbations in the measured densities of different species, which 
could be 
due to gravity waves. These observations suggest that dynamics in the upper 
atmosphere could significantly influence the neutral densities of N$_2$ 
and CO$_2$, and subsequently affect the NO density in the Martian upper 
atmosphere.

\cite{Stevens19} were able to analyze the IUVS/MAVEN observed
NO $\gamma$ limb intensity profiles for a small  period of observation i.e., 
6--8 
April 2016.  Based on the observed limb intensity, the
NO number densities were retrieved from the observed spectra 
between the altitudes 80 to 128 km, above which the 
retrieval 
becomes very difficult due to the presence of intense CO Cameron band emissions.
But our method of calculation is able to determine the NO 
number density 
profile from 120 to 200 km.   {Since the derivation of NO density  at higher 
altitudes  is difficult, we
suggest that our approach can be used as a 
baseline to constrain the NO abundance in the sunlit Martian
upper atmosphere for different MAVEN observational conditions. However,  the 
MAVEN observations are limited up to periapsis altitude  and it is 
difficult to compute the NO densities lower than 120 km altitude using this approach.}

\cite{Stevens19} modelled  NO density for the IUVS/MAVEN observation period and 
found 
that their derived density is smaller by 
a factor of 5 compared to the earlier Viking observations for the altitudes 
below 100 km. They ascribed this discrepancy to the assumed collisional rate 
coefficient for the reaction between
N($^4$S) and CO$_2$, which is 1 $\times$ 10$^{-18}$ cm$^3$ s$^{-1}$ instead of 
earlier \cite{Fox04} assumed 
value of 1 $\times$ 10$^{-16}$ cm$^3$ s$^{-1}$. 
By using the rate coefficient as used by \cite{Fox04}, our 
calculations in Figure~\ref{fig:pl_nnno} show that the contribution 
from N($^4$S) and CO$_2$ collisions is negligible to the total formation of NO  
for the altitudes above 120 km. By considering \cite{Stevens19} 
assumed rate coefficient in our calculations, which is two orders of magnitude smaller 
compared {to} the 
value used in the present work, we find no change in the modelled NO 
density profile. Thus, this 
calculation suggests that the contribution of N($^4$S) and CO$_2$ collisional
 reaction in the formation of NO  can be 
neglected  for the altitudes above 120 km.

Besides the observations of major neutral species, NGIMS/MAVEN also
provides NO density measurements in the Martian upper atmosphere. But various observations 
show that the NGIMS/MAVEN measured NO densities are higher by an order of 
magnitude 
compared to IUVS/MAVEN derived values. This  higher density is attributed to 
the recombination of N and O atoms inside the 
walls of mass spectrometer which produces additional NO. 
As \cite{Stevens19} noticed, 
the effect of contamination in  measuring the NO number density 
due to recombination of atomic oxygen and nitrogen cannot be 
neglected. By studying various factors associated in the calculation of NO 
density, \cite{Fox04} concluded that Viking 2 measurements
may not be accurate. Hence, the instrumental bias during Viking 2 measurements 
could be the main reason for the discrepancy between the modelled and 
observation of NO densities. Thus, there remains a compelling need of new 
measurements of NO densities in the Martian upper atmosphere to 
reconcile this long-standing problem.

The high emission rate factor makes NO (1,0) transition
 as the brightest feature of $\gamma$  band emission  at the 
wavelength 214.9 nm 
compared to other  transitions in the ultraviolet region. 
But this band emission is strongly obscured by CO (0,1) Cameron band emission 
(215.5 nm) due to 
proximity in the wavelength  and also due to its intensity. 
Using the derived NO densities, our
calculated limb intensities during deep dip 8 and 9 campaigns 
are consistent with the mean value 
of IUVS/MAVEN measurement  around 120 km (see Figure~\ref{fig:iuvs_no_cmp}). 
However, it should be cautioned that the uncertainty in the measured limb
radiance is large for the altitudes above 120 km \citep{Stevens19}.

\cite{Fox04} has modelled NO number density for low 
solar activity 
condition, by assuming a downward flux of -2 $\times$ 10$^7$
cm$^{-2}$ s$^{-1}$ at lower boundary i.e., 80 km.
\cite{Stevens19} have  reduced this downward flux value 
by two orders of magnitude to fit the IUVS/MAVEN retrieved NO
density profiles. They also noticed that assumed transport 
flux can affect
the shape of fitted profile only for altitudes below 100 km. 
But our calculated NO number densities around 120 km altitude are consistent 
with IUVS modelled 
profiles also
supports the assumption that the transport has a minor 
role in determining NO number density in the altitude range 120--160 km (see 
Figure~\ref{fig:no_cmp}). Hence, 
we suggest that the
NGIMS/MAVEN measured neutral and ion densities are suitable to
study NO density and also its dayglow emission intensity 
for different  seasonal conditions in the dayside upper
atmosphere of Mars in this altitude range.

As earlier discussed, the intense CO Cameron band inhibits 
the measurement of NO emission intensity in the dayglow
spectra. Our calculation in Figure~\ref{fig:iuvs_no_cmp} 
suggests that the emission intensity falls by an order of magnitude 
beyond 150 km, hence it is difficult to extract NO number density 
from the faint NO $\gamma$ band emission in the {background} of
intense CO Cameron band emission. Hence, we suggest that
the derivation of NO number density based on photochemistry and
subsequently calculating its dayglow emission intensity can 
serve as a baseline while analysing the ongoing and upcoming
IUVS/MAVEN observations. More such observations of NO dayglow emissions
along with modelling studies are necessary  to constrain its variability  in 
the Martian upper atmosphere.

\section{Summary and Conclusions}
\label{sec:sum_con}
The in-situ NO density measurements are 
difficult to make in the Martian upper atmosphere because of its low abundance
(more than 3 orders of magnitude compared to CO$_2$) and also highly reactive 
in nature. Moreover, the 
thermal recombination of atomic 
nitrogen and atomic oxygen inside  the mass spectrometer also inhibits the  
accurate
determination of NO density. Measurement of NO density from the remote dayglow 
observations is also difficult since its 
strongest ultraviolet emission feature, i.e., (1,0) gamma band, is severely 
obscured  by the 
intense CO Cameron band (50 times 
higher in magnitude). By accounting for major chemical reactions and using the 
NGIMS/MAVEN measured neutral and ion densities, we developed a 
photochemical model to study the photochemistry of NO in the Martian upper 
atmosphere.
MAVEN's deep dip campaigns provide a 
unique opportunity to study NO photochemistry in the dayside 
Martian upper atmosphere at the altitudes as low as 120 km. By utilizing the
NGIMS/MAVEN measured neutral and ion densities during MAVEN deep dip 8 and 9 
campaigns in our model, we 
present a method to calculate NO number density in the dayside 
upper atmosphere of Mars. Using the modelled NO  number density profiles we 
also calculated NO dayglow (1,0) $\gamma$ band emission 
intensity. 
Our calculated NO number density profiles are consistent with the IUVS/MAVEN
retrieved profile and also with the modelled values for IUVS observation conditions.
We found that the calculated NO number density varies by a
factor of 2--5 and consequently its dayglow intensity
due to the variation in  CO$_2$ and N$_2$ densities at
around 120 km. Hence, we suggest that the present method of calculation serves 
as a baseline to 
estimate the NO number density and also its $\gamma$ bands emission intensity
under  different seasonal and solar 
conditions in the dayside Martian upper atmosphere.
 {Based on our  calculations we suggest that  future
	MAVEN deep dip and remote  IUVS/MAVEN observations should be focused in the altitude 
	region 120 and 130 km  to constrain the NO density in the Martian upper atmosphere.}
More 
observations of NO emissions along with modelling studies are necessary 
to fully comprehend the NO density distribution on the 
dayside upper atmosphere of Mars.

\section*{Acknowledgements}
SR is supported by Department of Science 
and Technology (DST) with Innovation in Science Pursuit for 
Inspired Research (INSPIRE) faculty award 
[grant:dst/inspire/04/2016/002687], and he would like to 
thank Physical Research Laboratory for facilitating 
conducive research environment.  He is a visiting researcher at Khalifa Univeristy during a part of this work.
AB was J.C. Bose Fellow duirng the period of this work. 
A part of this work was done when MD was PDF at PRL and presently working 
at Maulana Azad National Institute of Technology Bhopal as an assistant professor.

\section*{Data Availability}
This paper make use of NGIMS/MAVEN measured 
neutral and ion number densities L2 data for deep dip 8
and 9 campaigns which has been accessed through the web link 
https://pds-atmospheres.nmsu.edu. The derived data generated
in this research will be shared on reasonable request
to the corresponding author.

\end{document}